\documentclass[letterpaper,aps,pre,twocolumn,showcaps,superscriptaddress,showpacs,amsmath,amssymb]{revtex4}
\usepackage{graphicx}

\newcommand{\dd}{\mathrm{d}}

\newcommand{\pd}[2]{\frac{\partial #1}{\partial #2}}
\newcommand{\fd}[2]{\frac{\delta #1}{\delta #2}}
\newcommand{\mean}[1]{\langle #1 \rangle}
\newcommand{\Int}[1]{\int\dd #1\;}
\newcommand{\IInt}[3]{\int_{#2}^{#3}\dd #1\;}

\renewcommand{\vec}[1]{\mathbf #1}

\newcommand{\mat}[1]{\mathsf #1}

\newcommand{\x}{\vec r}         
\newcommand{\shr}{\dot{\gamma}} 
\newcommand{\nois}{\boldsymbol\xi}
\newcommand{\Tk}{\theta}
\newcommand{\ps}{\psi_\text{s}}
\newcommand{\id}{\mathbf 1}

\newcommand{\T}{\text{T}}

\newcommand{\al}{\alpha}

\newcommand{\eps}{\varepsilon}

\newcommand{\om}{\omega}
\newcommand{\ra}{\rightarrow}

\begin{document}


\title{Effective Confinement as Origin of the Equivalence of Kinetic
  Temperature and Fluctuation-Dissipation Ratio in a Dense Shear Driven
  Suspension}

\author{Boris Lander}
\author{Udo Seifert}
\affiliation{{II.} Institut f\"ur Theoretische Physik,
  Universit\"at Stuttgart, Pfaffenwaldring 57, 70550 Stuttgart,
  Germany}
\author{Thomas Speck}
\affiliation{Institut f\"ur Theoretische Physik II,
  Heinrich-Heine-Universit\"at D\"usseldorf, Universit\"atsstra\ss e 1, 40225
  D\"usseldorf, Germany}

\begin{abstract}
  We study response and velocity autocorrelation functions for a tagged
  particle in a shear driven suspension governed by underdamped stochastic
  dynamics. We follow the idea of an effective confinement in dense
  suspensions and exploit a time-scale separation between particle
  reorganization and vibrational motion. This allows us to approximately
  derive the fluctuation-dissipation theorem in a ``hybrid'' form involving
  the kinetic temperature as an effective temperature \textit{and} an additive
  correction term. We show numerically that even in a moderately dense
  suspension the latter is negligible. We discuss similarities and differences
  with a simple toy model, a single trapped particle in shear flow.
\end{abstract}

\pacs{82.70.-y, 05.40.-a}

\maketitle


\section{Introduction}

Equilibrium statistical mechanics describes the connection of a few
macroscopic intensive quantities, e.g. pressure and temperature, with the
microscopic properties of the many particles constituting the
system~\cite{mcquarrie}. Although there is no equivalent formalism for systems
driven out of thermal equilibrium studies are often motivated by the quest for
simple governing principles such as an effective
temperature~\cite{cugl97a,cugl11}.  Of particular interest are small
mesoscopic systems such as colloidal particles, nanoparticles in solution, or
biological systems, all of which are dominated by fluctuations.

For systems only slightly perturbed from equilibrium -- into what is called
the \emph{linear response regime} -- the fluctuation-dissipation theorem (FDT)
relates the response with equilibrium correlations through the
temperature~\cite{kubo}. This unique temperature corresponds to what we would
measure with a thermometer, and is independent of the observables entering the
fluctuation-dissipation theorem. Examples for these observables are the
velocity related to the diffusion coefficient, or stress fluctuations
determining the viscosity. The practical importance of the FDT both for
experiments and simulations thus stems from the possibility to extract
transport properties from stationary fluctuations. In a more recent
application the FDT has been used to predict the self-assembly of model
systems~\cite{jack07,klot11}.

Even for systems driven into a non-equilibrium steady state (NESS) far beyond
the linear response regime we can still define a linear response
function~\cite{agar72,hang82,cris03,marc08}. For the broad class of driven
systems with Markovian dynamics and embedded in a fluid at well defined
temperature it has been demonstrated recently that then the FDT is most
convincingly interpreted in terms of an \emph{excess}
correlation~\cite{spec06,blic07,spec09,baie09,pros09,seif10,spec10}. The
single temperature that enters these generalized FDTs is that of the fluid,
and no approximations are involved. Nevertheless, for the purpose of a simple
description we might still be interested in defining an approximate
temperature through the fluctuation-dissipation ratio (FDR), i.e., the ratio
of correlation to response function. This strategy has originally been
proposed in the context of aging mean-field spin
systems~\cite{cugl97a,cugl11}, and subsequently been applied to many different
systems~\cite{ono02,haya04,haxt07}. In particular, in a sheared colloidal
suspension or fluid the Einstein relation between the self-diffusion
coefficient of a tagged particle and its mobility is broken and can be used to
define an effective temperature~\cite{bert02,szam04,krug09,land10,zhan11}.

\begin{figure}[b!]
  \centering
  \includegraphics[width=.9\linewidth]{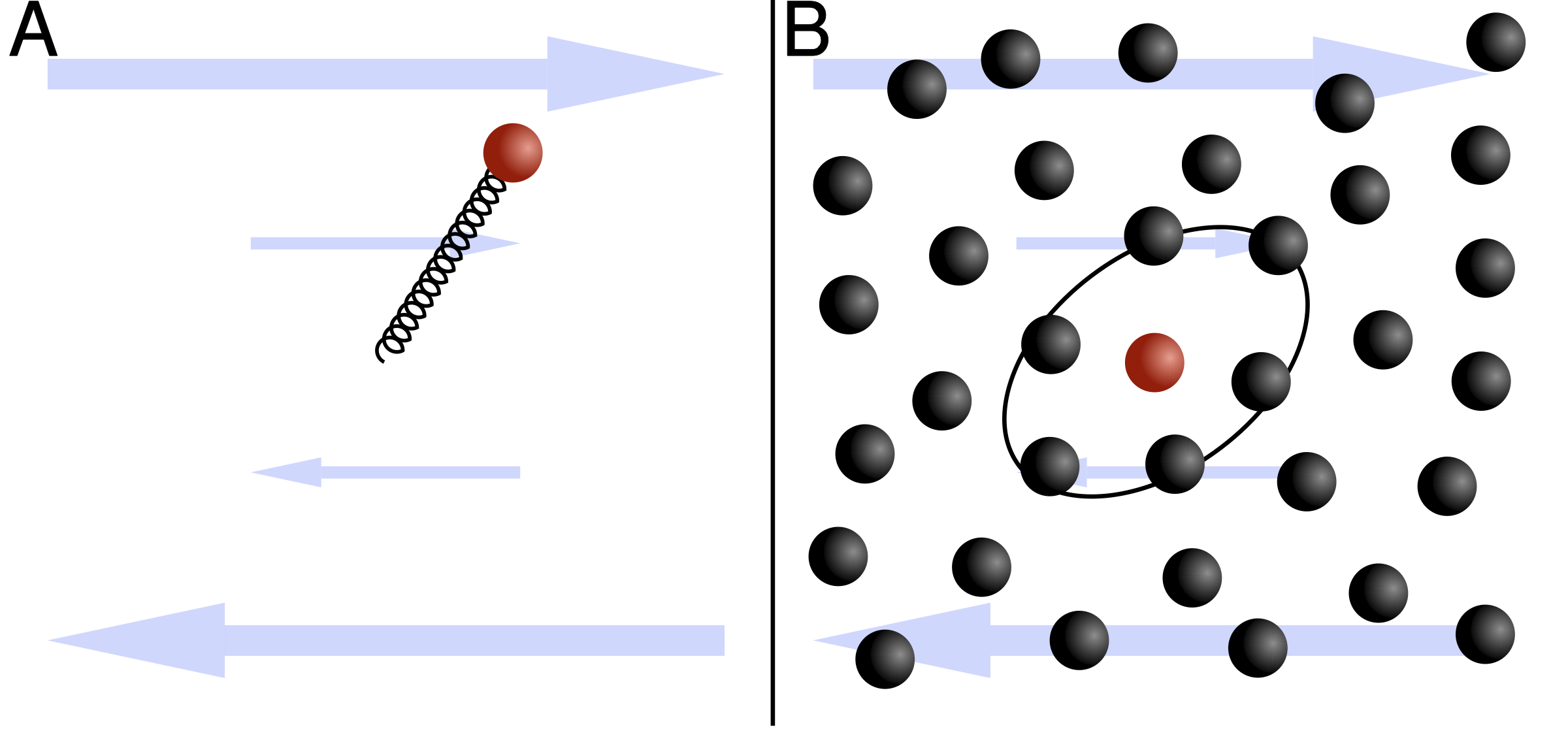}
  \caption{(Color online) Systems studied: (A)~Toy model with a single
    particle bound to the origin and (B)~tagged particle in a dense
    suspension. Both systems are driven through linear shear flow.}
  \label{fig:models}
\end{figure}

In this paper we study a many-body system governed by underdamped stochastic
dynamics and driven into a NESS through linear shear flow. Such a system could
model a fluid with every particle coupled to a stochastic thermostat,
colloidal or nano-suspensions, or dusty plasmas~\cite{shuk01}. We follow a
tagged particle, i.e., a randomly chosen particle out of many identical
interacting particles. Motivated by the physical picture of an effective
confinement in dense systems we also consider a single trapped colloidal
particle in shear flow~\cite{zieh09} as a toy model, see
Fig.~\ref{fig:models}. We discuss the FDT in a ``hybrid'' form: we relate
response and correlations through the kinetic temperature in the spirit of an
effective temperature, but with an additive correction term still present. For
the trapped particle expressions can be obtained analytically. For the tagged
particle we derive a similar FDT exploiting a time-scale separation due to the
effective confinement. In both cases we show that the correction term indeed
becomes negligible for strong confinement.


\section{Fluctuation-dissipation ratio}

We study the response of a single particle with mass $M$ moving in a viscous
liquid at temperature $T$. After applying a small external force $\vec f$
directly to the particle its mean velocity evolves as
\begin{equation}
  \label{eq:v:R}
  \mean{\vec v(t)} = \IInt{t'}{0}{t} \mat R(t-t')\vec f(t') + \mathcal O(f^2),
\end{equation}
where $\mat R=(R_{ij})$ is the response matrix with components
\begin{equation}
  \label{eq:R}
  R_{ij}(t-t') \equiv \left.\fd{\mean{v_i(t)}}{f_j(t')}\right|_{\vec f=0}
\end{equation}
with $t\geqslant t'$ due to causality. The brackets $\mean{\cdots}$ denote the
thermal average in the perturbed system. Throughout this paper we employ
dimensionless units and measure length in units of the particle diameter $a$
and energy in units of $k_\text{B} T$. Time is measured in units of
$\tau_0\equiv a^2/D_0$, which quantifies the time it takes for a free particle
with diffusion coefficient $D_0$ to diffuse a distance equal to its
diameter. The reduced mass $m\equiv(MD_0/k_\text{B} T)/\tau_0$ is the ratio of
the momentum relaxation time to the diffusive time scale $\tau_0$.

We describe the stochastic particle motion through the coupled equations
$\dot\x=\vec v$ and
\begin{equation}
  \label{eq:lang}
  m\dot{\vec v} = -\nabla U + \vec f - [\vec v-\vec u(\x)] + \nois,
\end{equation}
where $\x$ and $\vec v$ are the particle position and velocity,
respectively. Besides the conservative forces arising from the potential $U$
we can perturb the particle by a direct force $\vec f$. The noise $\nois$
modeling the interactions of the particle with solvent molecules has zero mean
and correlations
\begin{equation}
  \label{eq:nois}
  \mean{\nois(t)\nois^\T(t')} = 2\;\id\delta(t-t').
\end{equation}
The external shear flow enters through the term $\vec u(\x)=\shr y\vec e_x$,
i.e., the flow points in $x$-direction and increases its amplitude linearly
with the $y$-coordinate. Here, $\shr$ is the strain rate which in our units is
equal to the P\'eclet number.

Eq.~\eqref{eq:R} measures the linear response of the system. If the
unperturbed system ($\vec f=0$) is in thermal equilibrium the FDT
\begin{equation}
  \label{eq:fdt}
  \mat R(t-t') = \mat C(t-t') \equiv \mean{\vec v(t)\vec v^\T(t')}_0
\end{equation}
relates this response to the velocity auto correlation function (VACF) $\mat
C(t)$, where the subscript indicates that these correlations are to be
measured in the unperturbed system. 

For a computationally convenient representation of the response function
Eq.~\eqref{eq:R} valid both in and out of equilibrium consider the path weight
of the noise
\begin{equation}
  P[\nois(t)] \sim \exp\left\{ -\frac{1}{4}\Int{t} [\nois(t)]^2 \right\}.
\end{equation}
The stochastic velocity of the particle is a result of previous collisions
with solvent molecules, $\vec v(t)=\vec v[t;\nois(\tau)]$. Since a force
perturbation is equivalent to perturbing the noise we can write
\begin{equation}
  R_{ij}(t-t') =
  \int[\nois(\tau)] \fd{v_i[t;\nois(\tau)]}{\xi_j(t')}P[\nois(\tau)]
\end{equation}
for the response. A functional integration by parts then leads
to~\cite{spec06,cala05}
\begin{equation}
  \label{eq:R:nois}
  \mat R(t-t') = \frac{1}{2}\mean{\vec v(t)\nois^\T(t')}_0.
\end{equation}
Hence, even if the system is driven into a NESS the response can be measured
through a steady-state correlation function (see also
Refs.~\cite{chat04,bert07}). However, the FDT in the form Eq.~\eqref{eq:fdt}
no longer holds and the dimensionless fluctuation-dissipation ratio (FDR) is
defined as
\begin{equation}
  \label{eq:fdr}
  X_i(t) \equiv \frac{C_{ii}(t)}{R_{ii}(t)}
\end{equation}
for the diagonal components.

To calculate $X_i(0)$ we perform a short-time expansion of Eq.~\eqref{eq:v:R}
leading to $\mean{\vec v}\approx\Delta t\mat R(0)\vec f$. From the equation of
motion~\eqref{eq:lang} we obtain $m\mean{\dot{\vec v}}_0\approx m\mean{\vec
  v}/\Delta t=\vec f$. We have exploited that the average force on the
particle right before the perturbation vanishes, $\mean{\vec{\nabla}U}_0=0$,
which is quite obvious for isotropic systems but due to the inversion symmetry
about the origin it holds also in the presence of simple shear flow. Hence, in
our dimensionless units we obtain $R_{ij}(0)=\delta_{ij}/m$ and finally
\begin{equation}
  \label{eq:Tk}
  X_i(0) = m\mean{v_i^2}_0 \equiv \Tk_i.
\end{equation}
The right hand side is the \emph{kinetic temperature} as measured through the
velocity fluctuations. In the following we study for two systems whether, and
under which conditions, Eq.~\eqref{eq:Tk} extends to times $t>0$, i.e.,
whether $X_i(t)\approx\Tk_i$.


\section{Trapped particle}

The toy model we investigate first is a single particle trapped in the
harmonic potential
\begin{equation}
  \label{eq:harm}
  U(r) = \frac{1}{2}k r^2
\end{equation}
with strength $k$, where $\x$ is the displacement from the origin and
$r=|\x|$. Due to the linearity of the restoring force the $z$-component in
Eq.~\eqref{eq:lang} decouples and remains in equilibrium,
$X_z(t)=\Tk_z=1$. Therefore, in this section we only consider the motion in
the $xy$-plane.

\subsection{Analytical results}

Due to the quadratic potential Eq.~\eqref{eq:harm} the equations
of motion comprise a linear system of first order differential
equations. Hence, we can solve it for the velocity
\begin{equation}
  \label{eq:harm:v}
  \vec v(t) = \mat G^{vr}(t)\x_0 + \mat G^{vv}(t)\vec v_0
  + \frac{1}{m}\IInt{t'}{0}{t}\mat G^{vv}(t-t')\nois(t')
\end{equation}
given the initial displacement $\x_0$ and initial velocity $\vec v_0$. The
explicit expressions for the Green's functions $\mat G(t)$ are given in the
appendix~\ref{sec:green}. Both the VACF and the response function are easily
calculated from the solution Eq.~\eqref{eq:harm:v}. For the VACF we find
\begin{equation}
  \label{eq:harm:C}
  \mat C(t) = \mean{\vec{v}(t)\vec{v}_0^\T}_0 =
  \mat G^{vr}(t)\mean{\vec{r}_0\vec{v}_0^\T}_0 + \mat G^{vv}(t)
  \mean{\vec{v}_0\vec{v}_0^\T}_0,
\end{equation}
while the response function is trivially related to the Green's function
through
\begin{equation}
  \label{eq:harm:R}
  \mat R(t-t') = \frac{1}{2}\mean{\vec v(t)\nois^\T(t')}_0
  = \frac{1}{m}\mat G^{vv}(t-t')
\end{equation}
using the noise correlations Eq.~\eqref{eq:nois}. From the steady state
distribution Eq.~\eqref{eq:sim:psi} we obtain the moments
\begin{gather}
  \mean{\x_0\vec v_0^\T}_0 = \frac{1}{2k}\left(
    \begin{array}{cc}
      0 & -\shr \\ \shr & 0
    \end{array}\right),
  \\
  \mean{\vec v_0\vec v_0^\T}_0 = \frac{1}{m}\id
  + \frac{1}{2k}\left(
    \begin{array}{cc}
      \shr^2 & 0 \\ 0 & 0
    \end{array}\right).
\end{gather}
The kinetic temperatures Eq.~\eqref{eq:Tk} perpendicular to the shear flow are
$\Tk_y=\Tk_z=1$, whereas
\begin{equation}
  \label{eq:harm:Tk}
  \Tk_x = 1 + \al_x\shr^2 \geqslant 1, \qquad \al_x \equiv \frac{m}{2k}.
\end{equation}
As a consequence of the linear forces and the symmetry of the shear flow the
excess compared to equilibrium is proportional to $\shr^2$.

\begin{figure}[t]
  \centering
  \includegraphics{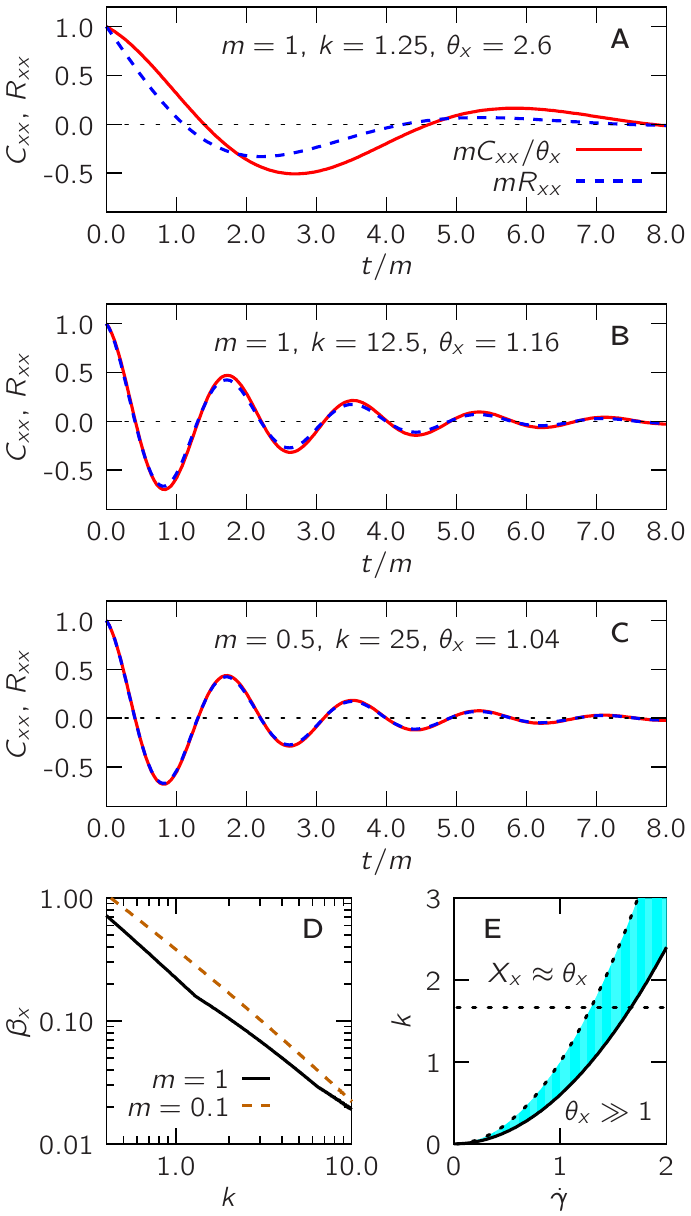}
  \caption{(Color online) Single particle moving in a harmonic trap: (A-C)
    scaled velocity auto-correlation function $C_{xx}(t)/\Tk_x$ and response
    function $R_{xx}(t)$ \textit{vs.} time $t$ for strain rate $\shr=2$ and
    different masses $m$ and trap strengths $k$. (B,C)~For large $k$, the
    correction in Eq.~\eqref{eq:harm:corr} becomes negligible and both curves
    lie on top of each other. (D)~The magnitude of the correction term
    $\beta_x$ as a function of $k$ for two different masses. (E)~Sketch of the
    different regimes of the FDT for $m=0.6$. The dashed lines $\shr^2$ and
    $1/m$ limit the region where the FDR is approximatly time-independent,
    $X_x\approx\Tk_x$. Below the solid line $m\shr^2$ the kinetic temperature
    is much larger than unity. While for the chosen $m$ there is a gap (shaded
    area), with increasing $m$ both regimes can be realized in the vicinity of
    the solid line.}
  \label{fig:harm}
\end{figure}

\subsection{The FDT}

Using the explicit expressions for the moments we see that with
$G^{vr}_{yx}(t)=0$ the correlation function for the $y$-component reads
$C_{yy}(t)=G^{vv}_{xx}\mean{v_y^2}_0$ and therefore $X_y(t)=\Tk_y=1$ at all
times. On the other hand, in the direction of the shear flow we obtain
\begin{equation}
  \label{eq:harm:corr}
  C_{xx}(t) = \Tk_xR_{xx}(t) + \mean{yv_x}_0 G^{vr}_{xy}(t),
\end{equation}
i.e., the velocity correlations are expressed through the response times the
kinetic temperature plus a correction term. Separating the dependence on
strain rate the correction term can be rewritten as
\begin{equation}
  \label{eq:beta}
  \mean{yv_x}_0 G^{vr}_{xy}(t) = \shr^2\beta_x I_x(t),
\end{equation}
where $\max|I_x(t)|=1$ and $\beta_x$ captures the magnitude of the correction
term. In Fig.~\ref{fig:harm}, we plot scaled response and correlation
functions for three representative values of the parameters $m$ and $k$. As
demonstrated in Fig.~\ref{fig:harm}D increasing the trap strength $k$ strongly
decreases $\beta_x$. The behavior of the FDT is sketched in
Fig.~\ref{fig:harm}E, where we compare the magnitude of the correction term to
the kinetic temperature. From Eq.~\eqref{eq:harm:Tk} we see that for $k\ll
m\shr^2$ we have $\Tk_x\gg 1$ (see also Fig.~\ref{fig:harm}A). For the
relevant case $k\gtrsim 1/m$ we find from the explicit expressions that for
the FDR to become approximately time-independent $k\gg\max\{\shr^2,m\shr^2\}$
must hold (Fig.~\ref{fig:harm}C). While for small masses $m<1$ there is a gap
(this case is sketched in Fig.~\ref{fig:harm}E), for sufficiently large
$m\gtrsim1$ these two regimes come close. In Fig.~\ref{fig:harm}B, we
demonstrate that there is indeed a regime of intermediate trap strength $k\sim
m\shr^2$ with an increased effective temperature where nevertheless
$X_x(t)\approx\theta_x$ holds to a very good degree.


\section{Tagged particle in a suspension}

The main system we now study is a suspension composed of $N$ particles in
which we tag and follow a single particle, say $k=1$, with position $\x_1$ and
velocity $\vec v_1$. The particles interact through a pair potential $u(r)$
with total potential energy $U=\sum_{i<j}u(|\x_i-\x_j|)$. Advection through
the shear flow leads to the well-known Taylor
dispersion~\cite{tayl53}. Instead of the absolute velocity $\vec v_1$ it will
be more convenient to use the relative velocity
\begin{equation}
  \vec v \equiv \vec v_1 - \vec u(\x_1)
\end{equation}
with respect to the flow as the observable entering the response $\mat R(t)$
and correlation function $\mat C(t)$.

\begin{figure*}[t]
  \centering
  \includegraphics{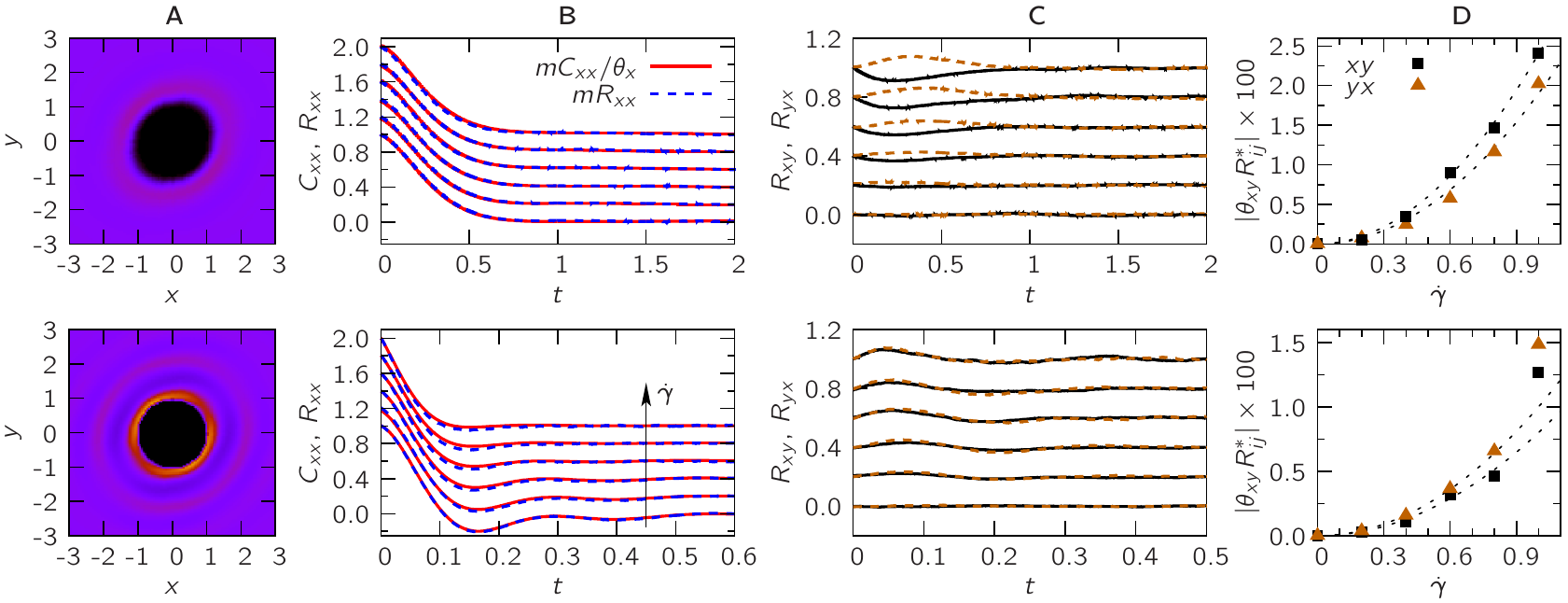}
  \caption{(Color online) Tagged particle with $m=1$ in a sheared suspension
    at volume fractions $\phi=0.1$ (top row) and $\phi=0.4$ (bottom row):
    (A)~Pair distribution function in the $xy$-plane for strain rate
    $\shr=1$. (B)~Scaled velocity auto-correlation functions $C_{xx}(t)/\Tk_x$
    and response functions $R_{xx}(t)$ for different strain rates: from bottom
    to top $\shr=0,0.2,0.4,0.6,0.8,1$. For visibility curves are
    shifted. (C)~Off-diagonal response functions $R_{xy}(t)$ (solid lines) and
    $R_{yx}(t)$ (dashed lines) for the same strain rates and offsets. (D)~The
    magnitude of the correction term $|\Tk_{xy}R^\ast_{ij}|$, i.e., the
    product of Eq.~\eqref{eq:Tk:off} with the maximum $R^\ast_{ij}$ of the
    off-diagonal response function, as a function of strain rate. The dashed
    lines show the quadratic fits.}
  \label{fig:susp}
\end{figure*}

\subsection{Time-scale separation}

In the following we assume a time scale separation between the motion of the
potential energy minimum (or inherent state position~\cite{stil84})
$\x_\text{c}$, and the vibrational motion of the tagged particle around
$\x_\text{c}$. The physical picture is that particles vibrate in a ``cage'' of
surrounding particles and that local reorganization takes much longer than the
vibrational motion. Linearizing the force exerted by neighboring particles on
the tagged particle leads to
\begin{equation}
  -\nabla_1 U \approx -\mat k[\x_1(t)-\x_\text{c}(t)], \qquad
  k_{ij} \equiv \left.\pd{^2U}{r_i\partial r_j}\right|_{\x=\x_\text{c}}.
\end{equation}
We solve the resulting equations of motion leading to the same formal result
Eq.~\eqref{eq:harm:C} for the correlation function and Eq.~\eqref{eq:harm:R}
the response function. In principle the Green's function can be calculated but
we will not need its explicit form here. To see that the first term in
Eq.~\eqref{eq:harm:C} vanishes consider the projected probability
\begin{equation}
  \bar\ps(\x_1,\vec v) = \Int{\x_2\cdots\dd\x_N}\Int{\vec v_2\cdots\dd\vec
    v_N}\ps(\{\x_i\})
\end{equation}
of the tagged particle, where $\ps$ is the full stationary distribution. For a
homogeneous, translationally invariant system $\bar\ps$ cannot depend on the
position $\x_1$ and therefore $\mean{\x_1\vec v^\T}_0\sim\mean{\vec v}_0=0$
vanishes. Hence, we obtain
\begin{equation}
  \label{eq:susp:corr}
  C_{ii}(t) = \Tk_i R_{ii}(t) + m\sum_{j\neq i}\mean{v_iv_j}_0R_{ij}(t)
\end{equation}
as our central result. The correction term differs from
Eq.~\eqref{eq:harm:corr} and now couples to the off-diagonal elements of the
response function instead of the off-diagonal component of the Green's
function connecting velocity and position.

\subsection{Langevin dynamics simulations}

We perform Langevin dynamics simulations to study Eq.~\eqref{eq:susp:corr} for
a specific system. The $N=1728$ particles are enclosed in a cubic simulation
box with edge length $L$. The particles interact through the Yukawa (or
screened Coulomb) pair potential
\begin{equation}
  \label{eq:yuk}
  u(r) =
  \begin{cases}
    \eps\frac{e^{-\kappa(r-1)}}{r} & (r\geqslant 1) \\
    \infty & (r<1),
  \end{cases}
\end{equation}
where $\eps$ is the interaction energy at contact and $\kappa^{-1}$ the
screening length determined by the composition of the surrounding solvent. We
choose $\eps=8.0$ and $\kappa=5.0$ in order to obtain a broad range of
densities for which the liquid phase is stable~\cite{azha00}. For the shear
flow we employ Lees-Edwards boundary conditions, which enforce a linear
velocity profile in the suspension~\cite{allen}. We integrate the equations of
motion by a stochastic velocity Verlet algorithm~\cite{frenkel} with a time
step $5\times10^{-4}$. Since the hard-core repulsion cannot be implemented in
the interaction potential we employ a simple algorithm that detects collisions
and computes the appropriate positions and velocities after the impact
according to momentum and energy conservation (see Refs.~\cite{stra99,foss00}
and references therein). The NESSs are prepared by initializing the particle
positions on a regular lattice at low density. Then we equilibrate the system
and slowly increase the density by decreasing the box size. After this
equilibrium system is constructed we slowly ramp up the strain rate until the
target value $\shr$ is reached. We simulate another $1000$ time steps to relax
the system into the steady state. This procedure is repeated separately for
independent runs.

We study suspensions at different volume fractions $\phi\equiv\pi N/(6L^3)$
and strain rates $\shr$. To determine the response and correlation functions
we simulate the motion of the system in the NESS and record the velocity
trajectories of $200$ randomly chosen particles in four independent runs. From
this data we can easily evaluate the VACF. To compute the response to a small
force we again employ Eq.~\eqref{eq:R:nois}, which is still valid for the many
particle system. This enables us to obtain the response function from steady
state trajectories without the need to explicitly perturb the
system. Therefore, in addition to the velocity, we also record the stochastic
forces acting on the tagged particles, which are directly accessible in a
numerical simulation.

\subsection{Numerical results}

The pair distribution function quantifies the probability to find another
particle at a displacement $\x$. It is distorted from its isotropic
equilibrium shape, see Fig.~\ref{fig:susp}A. This function visualizes the
average environment of the tagged particle. Under the influence of the shear
flow there is a higher number of particles in the compressional zones and a
depletion in the extensional zones~\cite{foss00}. Moreover, the peak
corresponding to the first nearest-neighbor shell becomes more pronounced.

\begin{figure}[b!]
  \centering
  \includegraphics{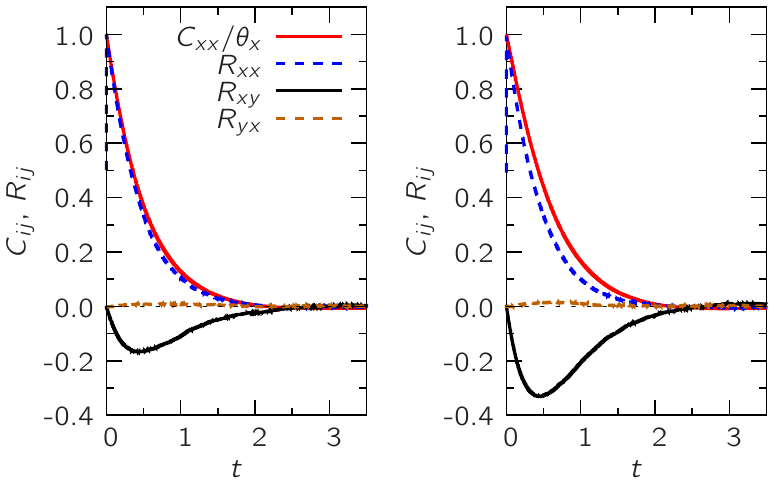}
  \caption{(Color online) Dilute suspension at $\phi=0.01$ and strain rate
    $\shr=1$ (left) and $\shr=2$ (right). In comparison to Fig.~\ref{fig:susp}
    a deviation between response $R_{xx}(t)$ and velocity autocorrelation
    function $C_{xx}(t)$ is observed.}
  \label{fig:low}
\end{figure}

In Fig.~\ref{fig:susp}B, the response functions $R_{xx}(t)$ together with the
scaled VACF $C_{xx}(t)/\Tk_x$ is plotted for volume fractions $\phi=0.1$ and
$\phi=0.4$, and for a range of strain rates. Even for times $t>0$ these
functions coincide, which implies that the additive correction term in
Eq.~\eqref{eq:susp:corr} vanishes. The off-diagonal response functions are
plotted in Fig.~\ref{fig:susp}C. For comparison response and correlation
functions are plotted in Fig.~\ref{fig:low} for a dilute suspension at volume
fraction $\phi=0.01$. Here we observe a clear difference between correlations
and response. To understand the observed behavior one should bear in mind that
we consider the velocities relative to the local flow. At low densities
collisions are rare and the particles adapt smoothly to the flow of the
solvent. Therefore, the largest deviations from the flow profile arise when a
particle diffuses in $y$-direction entering a region of faster or slower flow
in $x$-direction, to which it needs to adapt. For higher densities collisions
become much more frequent. These collisions prevent the particles from
adapting to the solvent flow. Additionally they distribute the momentum,
transfered to the particles by the shear flow, more or less randomly in the
three spacial directions. The result is that under shear flow the diagonal
velocity moments grow with increasing density, and become more and more
similar in size.

The two components $\mean{v_xv_z}_0\simeq0$ and $\mean{v_yv_z}_0\simeq0$ of
the velocity correlation matrix are very small (see also
Fig.~\ref{fig:mass}). Defining the off-diagonal ``temperature''
\begin{equation}
  \label{eq:Tk:off}
  \Tk_{xy} \equiv m\mean{v_xv_y}_0
\end{equation}
we see that the dominant contribution to the correction term in
Eq.~\eqref{eq:susp:corr} is $\Tk_{xy}R_{xy}(t)$ for the $x$ component, and
$\Tk_{xy}R_{yx}(t)$ for the $y$ component. In analogy to the trapped particle
[Eq.~\eqref{eq:beta}] we separate the strain rate dependence of the correction
terms,
\begin{equation}
  \Tk_{xy}R_{xy}(t) \approx \shr^2\beta_x I_x(t), \quad
  \Tk_{xy}R_{yx}(t) \approx \shr^2\beta_y I_y(t),
\end{equation}
with again $\max|I_i(t)|=1$ and coefficients $\beta_i$. In
Fig.~\ref{fig:susp}D, we plot $|\Tk_{xy}R^\ast_{ij}|\approx\shr^2\beta_i$,
where $R^\ast_{ij}$ is the maximal value of the off-diagonal component of the
response matrix. For $\phi=0.1$ we observe the predicted quadratic dependence
on the strain rate $\shr$. For $\phi=0.4$ the quadratic predicition holds for
$\shr\leqslant0.8$, while for larger strain rates higher order terms in $\shr$
become important.

\begin{figure}[t]
  \centering
  \includegraphics{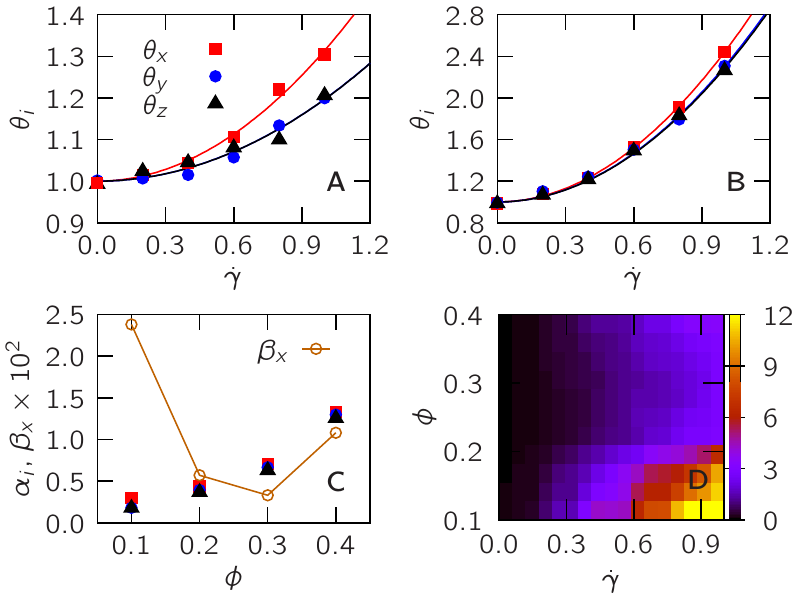}
  \caption{(Color online) The kinetic temperatures $\Tk_i$ (symbols) and their
    quadratic fits (lines) plotted \textit{vs}. strain rate $\shr$ for
    $m=1$. Shown are the three directions for volume fractions (A)~$\phi=0.1$
    and (B)~$\phi=0.4$. In (C) the coefficients $\alpha_i$ and
    $\beta_x\times10^2$ are plotted as function of the volume fraction
    $\phi$. (D)~Relative magnitude of the correction term
    $|\Tk_{xy}R^\ast_{xy}|/\Tk_x\times10^3$ as function of strain rate and
    density, cf. Fig.~\ref{fig:harm}E. Colors are linearly interpolated.}
  \label{fig:teff}
\end{figure}

In Fig.~\ref{fig:teff}, we plot the kinetic temperatures as a function of
strain rate. While for $\phi=0.1$ a clear difference between motion parallel
to the flow ($\Tk_x$) and motion perpendicular to the flow
($\Tk_y\simeq\Tk_z$) can be seen, this distinction is diminished at higher
densities. Moreover, all kinetic temperatures can be well fitted by the
quadratic function Eq.~\eqref{eq:harm:Tk} with coefficients $\al_i$ for the
three directions. The increase of velocity fluctuations can be explained by
forced collisions due to the flow gradient, an effect that is more pronouned
at higher densities and higher strain rates.

\begin{figure}[t]
  \centering
  \includegraphics[width=.9\linewidth]{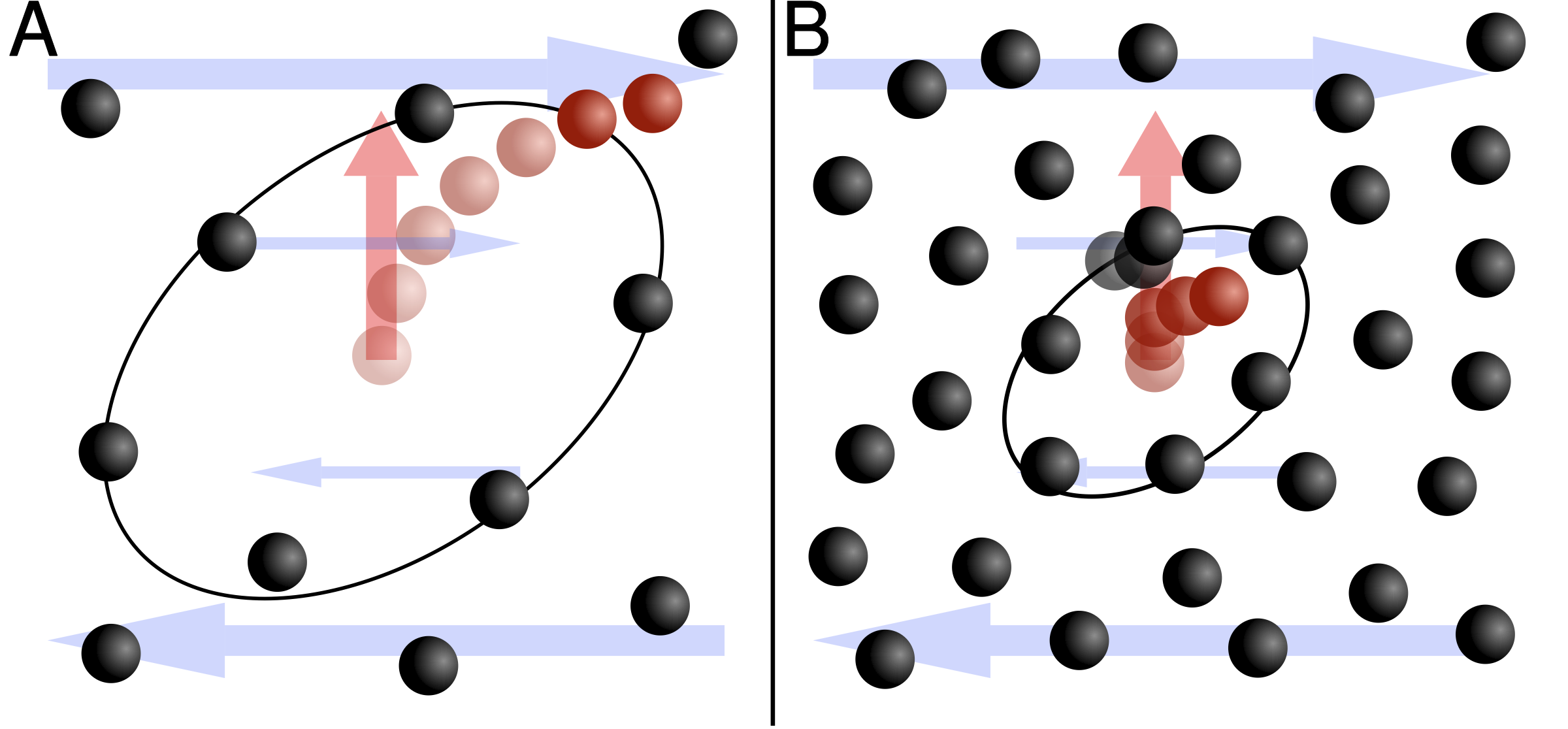}
  \caption{(Color online) Sketch of the time evolution after a force
    perturbation in $y$-direction (arrow) determining the shape of the
    $R_{xy}(t)$ response function: (A)~At low densities the tagged particle is
    slower than the surrounding flow field due to inertia. (B)~At high
    densities collisions with neighboring particles are more likely, pushing
    the tagged particle in the direction of the flow.}
  \label{fig:resp}
\end{figure}

In Fig.~\ref{fig:teff}C, the coefficients $\al_i$ and $\beta_x$ are shown for
the different densities. The coefficient $\beta_x$ decreases for larger
densities but then turns up again at $\phi=0.4$. The reason for this
non-monotonic behavior is that there are two effects determining the shape of
the $xy$-component of the response function, see Fig.~\ref{fig:resp}. One
dominates for lower, one for higher densities. At low densities forcing the
particle upwards in $y$-direction moves it into a region of faster flow. While
the velocity relaxes the relative velocity with respect to the shear flow is
negative. At higher densities this effect is weaker because of more frequent
particle collisions, and thus large excursions in $y$-direction are
rare. Hence, the velocity differences due to the motion are smaller and so is
the response caused by this effect. In addition another effect of the
collisions becomes more significant. Pulling the particle upwards in
$y$-directions makes collisions with particles from the left more likely than
with particles from the right. This leads to an average acceleration to the
right which counteracts the first effect. In the intermediate density regime
these two effects almost cancel and lead to a small $\beta_x$.

Finally, in Fig.~\ref{fig:teff}D the magnitude of the correction term with
respect to the response function, $|\Tk_{xy}R^\ast_{xy}|/\Tk_x$, is plotted,
see also Fig.~\ref{fig:harm}E. Note that over the whole parameter range
studied here this value is lower than $0.02$, i.e., $X_x\approx\Tk_x$ holds to
a very good degree. However, we see that for large strain rates and low
density this ratio grows by an order of magnitude. Due to the effects
described above for the highest density the range of strain rates for which
the correction term is negligible shrinks again.

\subsection{Overdamped limit}

\begin{figure}[t]
  \centering
  \includegraphics{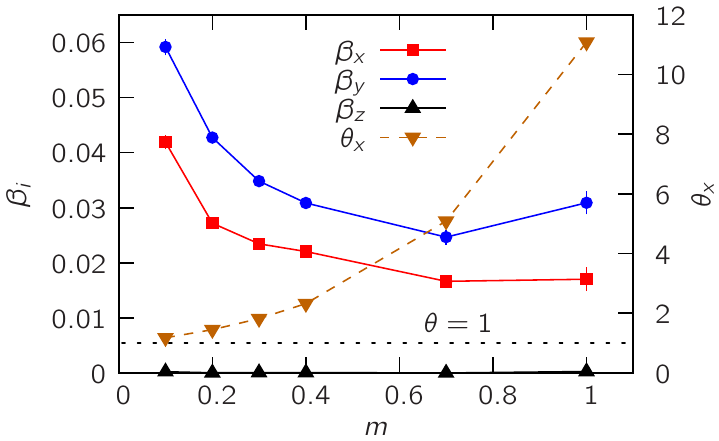}
  \caption{(Color online) The coefficients $\beta_i$ (solid lines, left axis)
    and the kinetic temperature $\theta_x$ (dashed line, right axis) as a
    function of the reduced mass $m$ for volume fraction $\phi=0.4$ and strain
    rate $\shr=2$.}
  \label{fig:mass}
\end{figure}

Our results have been derived for systems with underdamped stochastic
dynamics. Of greater practical importance in colloidal suspensions is the
overdamped limit corresponding to neglecting inertia, $m\ra0$. In
Fig.~\ref{fig:mass}, the dependence of the coefficients on the reduced mass $m$
is plotted. As expected for smaller $m$ the kinetic temperatures approach
unity, $\theta_i\approx1$. The additive correction term is no longer
negligible as $\beta_x$ and $\beta_y$ grow. We thus recover the
fluctuation-dissipation theorem as described in the introduction, in which the
bath temperature enters and the equilibrium form of the FDT is completed by an
excess correlation function.

Inserting the overdamped equation of motion for the tagged particle into
Eq.~\eqref{eq:R:nois} one arrives at
\begin{equation}
  \label{eq:ov}
  C_{ij}(t) = 2R_{ij}(t) + \mean{F^{(1)}_i(t)F^{(1)}_j(0)}_0
\end{equation}
for $t>0$ and components $i,j=y,z$ perpendicular to the shear
flow~\cite{land10}. Here, $\vec F^{(1)}=-\nabla_1U$ is the force on the tagged
particle exerted by its neighboring particles. The time-integrated version of
Eq.~\eqref{eq:ov} has been obtained previously within mode-coupling
calculations~\cite{szam04,krug09,krug10}. For systems in which the force-force
correlations can be neglected Eq.~\eqref{eq:ov} predicts a universal FDR
$X_i=2$. It has been argued that such an approximation might be justified in
dense suspensions close to the glass transition under sufficiently large
shear.

\section{Summary}

We have studied the relation between velocity autocorrelation and response
function of a tagged particle moving in a suspension that is driven into a
non-equilibrium steady state through simple shear flow. Under the assumption
of a time-scale separation between vibrational motion and local reorganization
the tagged particle effectively behaves like a trapped particle. The diagonal
components of the tagged particle's velocity autocorrelation function are then
given by
\begin{equation}
  C_{ii}(t;\shr) \approx \Tk_i(\shr)R_{ii}(t;\shr) + \shr^2\beta_iI_i(t;\shr)
\end{equation}
with expansion of the kinetic temperature
\begin{equation}
  \Tk_i(\shr) \approx 1+\al_i\shr^2.
\end{equation}
Here, $R_{ii}(t;\shr)$ are the response functions Eq.~\eqref{eq:R},
$I_i(t;\shr)$ are functions of order unity, and $\al_i$ and $\beta_i$ are
coefficients independent of the strain rate $\shr$. While these expressions
are exact for the trapped single particle our numerical results show that they
hold approximately to a very good degree for a tagged particle moving in an
interacting colloidal suspension. One might of course anticipate the quadratic
dependence on strain rate from symmetry arguments close to equilibrium. We
have shown here that these expressions follow from a time-scale separation
caused by an effective confinement.

The effect of the shear flow is to break symmetry and to couple the particle
velocity to earlier perturbations perpendicular to its velocity. This leads to
an additive correction term that grows with $\shr^2$. Since $\beta_i\ll1$ this
correction is negligible up to dimensionless strain rates
$\shr\sim(\max\{\beta_i\})^{-1/2}$, which can be far from
equilibrium. Specifically, here we have studied strain rates in the range
$\shr\leqslant1$ and found excellent agreement between correlation and
response functions, see Fig.~\ref{fig:susp}. However, already at
$\shr\gtrsim2$ the kinetic temperatures start to divert from the quadratic
law, indicating the importance of higher order terms. Increasing the density
the tagged particle interacts more strongly with its surrounding
particles. The kinetic temperature increases due to more frequent collisions
with neighboring particles in conjunction with transport due to the flow. This
is in contrast to the trapped particle where tightening the trap reduces
fluctuations and therefore the kinetic temperature approaches the bath
temperature.

An intriguing perspective is to apply our results to supercooled (or
supersaturated) conditions. Since dynamics slows down dramatically and the
time-scale separation between vibrations and long-lived particle displacements
becomes even more pronounced we expect that our results extend into the
supercooled regime. While we have employed stochastic dynamics one might
speculate that our results also hold in systems governed by deterministic
dynamics such as the SLLOD equations of motion~\cite{tuck97} as employed in
Ref.~\cite{bert02}. Future work will also address the influence of
hydrodynamic interactions and flow generated through boundaries.

\acknowledgments

We acknowledge financial support by the DFG through project SE~1119/3. TS
acknowledges funding through Alexander von Humboldt foundation and, during
early stages of the project, by the Director, Office of Science, Office of
Basic Energy Sciences, Materials Sciences and Engineering Division and
Chemical Sciences, Geosciences, and Biosciences Division of the
U.S. Department of Energy under Contract No.~DE-AC02-05CH11231.


\appendix
\section{Green's function for a particle in a harmonic trap in shear flow}
\label{sec:green}

The equations of motion~\eqref{eq:lang} for the potential Eq.~\eqref{eq:harm}
are most conveniently written as $\dot{\vec x}\equiv\mat A\vec x+(0,\nois)^\T$
for a vector $\vec x=(x,y,v_x,v_y)^\T$ with
\begin{equation*}
  \mat A = \frac{1}{m}\left(
    \begin{array}{cccc}
      0 & 0 & m & 0 \\
      0 & 0 & 0 & m \\
      -k & \shr & -1 & 0 \\
      0 & -k & 0 & -1
    \end{array}\right).
\end{equation*}
The Green's function is
\begin{equation*}
  \mat G(t) \equiv e^{\mat A t} = \left(
    \begin{array}{cc}
      \mat G^{rr}(t) & \mat G^{rv}(t) \\
      \mat G^{vr}(t) & \mat G^{vv}(t)
    \end{array}\right).
\end{equation*}
We need the explicit expressions for the following two matrices:
\begin{gather*}
  \mat G^{vr}(t) = e^{-t/2m}\left[-\left(\om+
      \frac{1}{4m^2\om}\right)\sin\om t\id + \shr g_{vr}(t)\id_{xy}
  \right], \\
  g_{vr}(t) \equiv \frac{(4m^2\om^2-1)\sin\om t+
    (4m^2\om^2+1)\om t\cos\om t}{8m^3\om^3}, \\
  \mat G^{vv}(t) = e^{-t/2m}\left[\left(\cos\om t-\frac{\sin\om
        t}{2m\om}\right)\id + \shr g_{vv}(t)\id_{xy} \right], \\
  g_{vv}(t) \equiv \frac{(2m\om^2 t-1)\sin\om t+\om t\cos\om
    t}{4m^2\om^3},
\end{gather*}
where
\begin{equation*}
  \om \equiv \frac{\sqrt{4km-1}}{2m}, \qquad
  \id_{xy} \equiv \left(\begin{array}{cc}
      0 & 1 \\ 0 & 0
    \end{array}\right).
\end{equation*}
The diagonal components are independent of the strain rate. To determine
$\beta_x$ as defined in Eq.~\eqref{eq:beta} we calculate numerically the
maximum of $G^{vr}_{xy}(t)$.

For the sake of completeness, the other two matrices are
\begin{gather*}
  \mat G^{rr}(t) = e^{-t/2m}\left[\left(\cos\om t+\frac{\sin\om
        t}{2m\om}\right)\id + \shr g_{rr}(t)\id_{xy} \right], \\
  g_{rr}(t) \equiv \frac{(2m\om^2t+1)\sin\om t-\om t\cos\om
    t}{4m^2\om^3}, \\
  \mat G^{rv}(t) = e^{-t/2m}\left[\frac{\sin\om t}{\om}\id
    + \shr g_{rv}(t)\id_{xy} \right], \\
  g_{rv}(t) \equiv \frac{\sin\om t-\om t\cos\om t}{2m\om^3}.
\end{gather*}
The stationary distribution
\begin{equation}
  \label{eq:sim:psi}
  \ps(\x,\vec v) = \frac{1}{(2\pi)^2\sqrt{\det\mat M}}\exp\left\{
    -\frac{1}{2}\vec x\cdot\mat M^{-1}\vec x \right\}
\end{equation}
is Gaussian and therefore determined by the symmetric covariance matrix
\begin{equation*}
  \mat M \equiv \left(
    \begin{array}{cc}
      \mean{\x\x^\T}_0 & \mean{\x\vec v^\T}_0 \\
      \mean{\vec v\x^\T}_0 & \mean{\vec v\vec v^\T}_0
    \end{array}\right)
\end{equation*}
alone. We calculate $\mat M$ using Chandrasekhar's theorem~\cite{dhont},
\begin{equation*}
  \mat M = \frac{2}{m^2}\IInt{t}{0}{\infty} \left(
    \begin{array}{cc}
      \mat G^{rv}(t)\mat G^{rv\T}(t) & \mat G^{rv}(t)\mat G^{vv\T}(t) \\
      \mat G^{vv}(t)\mat G^{rv\T}(t) & \mat G^{vv}(t)\mat G^{vv\T}(t)
    \end{array}\right).
\end{equation*}
With $1+4(m\om)^2=4km$, we obtain the stationary correlations
\begin{gather}
  \label{eq:mom:rr}
  \mean{\x\x^\T}_0 = \frac{1}{k}\id
  + \frac{1}{2k^2}\left(
    \begin{array}{cc}
     \frac{1+4km}{4k}\shr^2 & \shr \\
      \shr & 0
    \end{array}\right), \\
  \label{eq:mom:rv}
  \mean{\x\vec v^\T}_0 = \frac{1}{2k}\left(
    \begin{array}{cc}
      0 & -\shr \\ \shr & 0
    \end{array}\right), \\
  \label{eq:mom:vv}
  \mean{\vec v\vec v^\T}_0 = \frac{1}{m}\id
  + \frac{1}{2k}\left(
    \begin{array}{cc}
      \shr^2 & 0 \\ 0 & 0
    \end{array}\right).
\end{gather}
The distribution function projected into the configuration space is obtained
through integrating out the velocities in the stationary
distribution~\eqref{eq:sim:psi},
\begin{equation*}
  \begin{split}
    \Psi_\text{s}(\x) &= \Int{\vec v}\ps(\x,\vec v) \\
    &= \frac{1}{2\pi\sqrt{\det\mean{\x\x^\T}_0}}
    \exp\left\{-\frac{1}{2}\x\cdot\mean{\x\x^\T}_0^{-1}\x \right\}.
  \end{split}
\end{equation*}
In the overdamped limit one obtains
\begin{equation*}
  \mean{\x\x^\T}_0 \xrightarrow{m\ra0} \frac{1}{k} \left(
    \begin{array}{cc}
      1+2\bar\shr^2 & \bar\shr \\ \bar\shr & 1
    \end{array}\right)
\end{equation*}
with $\bar\shr\equiv\shr/2 k$.


\end{document}